\documentclass{elsart}[12pt]
\usepackage[dvips]{graphics}
\begin{document}
\begin{frontmatter}

\title{Kinetic energy spectra for fragments and break-up density in
multifragmentation}

\small
\author[IPNO,NIPNE]{Ad. R. Raduta\thanksref{corres}},
\author[IPNO]{B. Borderie},
\author[IPNO]{E. Bonnet},
\author[IPNO]{N. Le Neindre}, 
\author[FLO]{S. Piantelli} and
\author[IPNO]{M. F. Rivet}

%--------------------------------------------------------------------
\address[IPNO]{Institut de Physique Nucleaire, IN2P3-CNRS, F-91406 Orsay
cedex, France}
\address[NIPNE]{National Institute for Physics and Nuclear Engineering,
Bucharest-Magurele, POB-MG 6, Romania}
\address[FLO]{Dip. di Fisica e Sezione INFN, Universit\`a di Firenze, I-50019
Sesto Fiorentino (Fi), Italy}
%----------------------------------------------------------------------
\thanks[corres]{Corresponding author, araduta@ifin.nipne.ro}
%----------------------------------------------------------------------
\normalsize

\begin{abstract}
We investigate the possibility, in nuclear fragmentation, to extract
information on nuclear density at break-up from fragment kinetic energy spectra
using a simultaneous scenario for fragment emission. 
It is found that a decrease of peak centroids for kinetic energy spectra of fragments
with increasing excitation energy can be observed at constant low density,
which is different from recently published results of
Viola {\it et al.} \cite{viola}. 
%assuming a sequential fragment emission.

\end{abstract}
\begin{keyword}
Multifragmentation \sep fragment kinetic energy \sep  microcanonical model
\PACS 25.70.Pq \sep 24.10.Pa
\end{keyword}
\end{frontmatter}

One of the most challenging tasks of nuclear physics in the last
decades is the
determination of the phase diagram of excited atomic nuclei.
Despite the important theoretical
and experimental work already done, the problem is far from being solved. From
the experimental
point of view the localization of nuclear multifragmentation data in the phase
diagram requires accurate independent measurements
of temperature and density at the break-up stage.
While the problem of
temperature determination has been solved with acceptable accuracy up to 5-6
MeV \cite{poch,nato,dasg,bbor}, no satisfactory method
to determine the spatial extension of the presumably equilibrated nuclear
system at break-up has
been proposed. Thus, experiments using light particles
interferometry \cite{aladin} indicate
freeze-out densities ranging from less than $\rho_0/10$ to $\rho_0/2.5$ 
(where $\rho_0$ is the normal
nuclear density); on the other side, statistical \cite{smm,mmmc,deses,lbeaulieu,mmm} and
dynamical models \cite{bob,frankland} succeed 
to describe well the available experimental data with freeze-out densities in
the interval $\rho_0/9$ to $\rho_0/2.5$. 

Ref. \cite{viola} tries to obtain information on break-up density
using kinetic energy spectra of intermediate mass fragments
(IMF: $3 \leq Z \leq 15$) measured in light ion induced
multifragmentation of gold, namely $^4$He+$^{197}$Au at
50 MeV/nucleon \cite{zhang},
$^{14}$N+$^{197}$Au at 20-100 MeV/nucleon \cite{wile} and 
$^3$He+$^{197}$Au at 4.8 GeV bombarding energy \cite{bracken}.

The pattern of IMF kinetic energy spectra led the authors of Ref. \cite{viola}
to fit the extracted equilibrium sources with a Maxwell-Boltzmann type 
distribution,
\begin{equation}
\frac{dN}{dK}=(K-V'_C) \cdot \exp\left( - \frac{(K-V'_C)}{T_s}\right),
\label{eq:MB}
\end{equation}
where $K$ is the kinetic energy of the considered
cluster ($A_F$,$Z_F$) emitted by the source ($A_s$,$Z_s$),
$V'_C$ the cluster kinetic Coulomb energy and $V_C$ the
Coulomb barrier
between the emitted fragments and the residual nucleus,
\begin{equation}
V_C=1.44 \cdot \frac{Z_F(Z_s-Z_F)}{d \left( A_F^{1/3}+(A_s-A_F)^{1/3} \right )},
\label{eq:cb}
\end{equation}
\begin{equation}
V'_C=\frac{A_s-A_F}{A_s} \cdot V_C;
\label{eq:cbf}
\end{equation}

$T_s$ is the temperature of the multifragmenting source.

Thus, the interpretation of the behavior
of these spectra with the rise of excitation energy is made within a
%scenario
parameterization suitable for sequential particle emission. 
From Eq. (\ref{eq:MB}) results that a temperature increase will
determine a shift of the
centroids of the Coulomb-like peaks toward higher values of $K$ together
with the broadening of the distribution while a decrease
of the Coulomb barrier (by increasing the fragments' centre relative distances
expected at low density)
will shift the distribution in the opposite direction. Starting from these
premises Ref. \cite{viola} presents a systematic fit over an important
collection of experimental
spectra corresponding to an excitation energy interval ranging from 0.9
to 7.9 MeV/nucleon
and reaches the conclusion that the displacement of the maximum
of $dN/dK$ IMF distributions towards lower energy and
observed in the range 2-5 MeV/nucleon is a sufficient evidence
in favor of decreasing
break-up density down to $\sim \rho_0/3$ with increasing excitation energy.

Both the short time scale characterizing the decay of nuclei
with excitation energies exceeding
%2 MeV/nucleon
3 MeV/nucleon and the
pattern of fragments' relative velocities indicate that
multifragmentation
%is
should be treated as a simultaneous process
\cite{durand,beaulieu}. 
In this framework, do the displacements of peak centroids of
kinetic energy spectra 
%may
reveal a decrease of the nuclear break-up density?
We shall demonstrate in this letter that
%the expectation proposed in Ref. \cite{viola} within a sequential
%scenario is no more valid as far as a simultaneous process is involved.
such displacements are then obtained at constant low density.
%%%%%%%%%%%%%%%%%%%%%%%%%%%%%%%%%%%%%%%%%%%%%%%%%%%%%%%%%%%%%%%%%%%%
\begin{figure}
\resizebox{0.99\textwidth}{!}{%
  \includegraphics{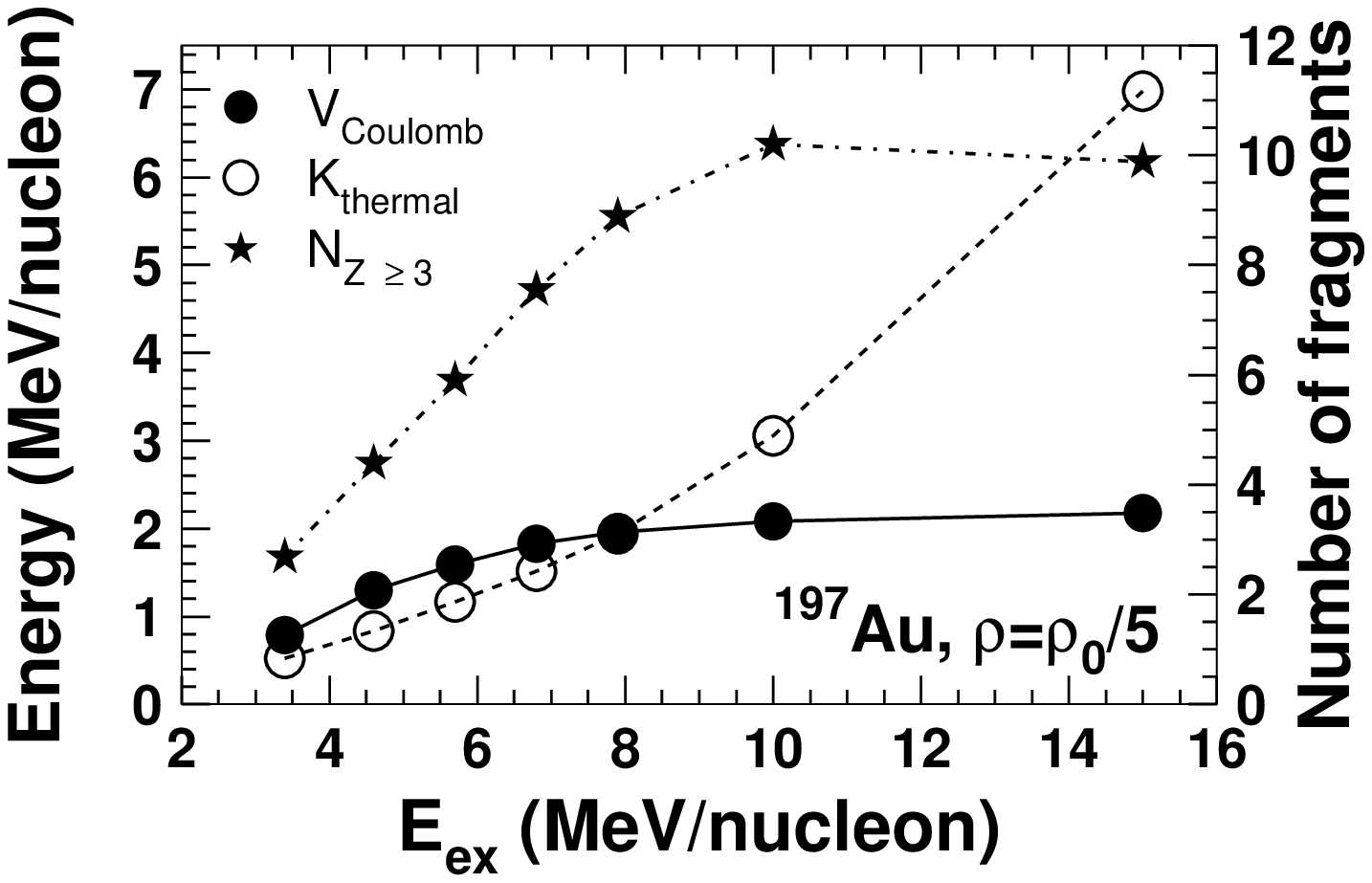}
  \includegraphics{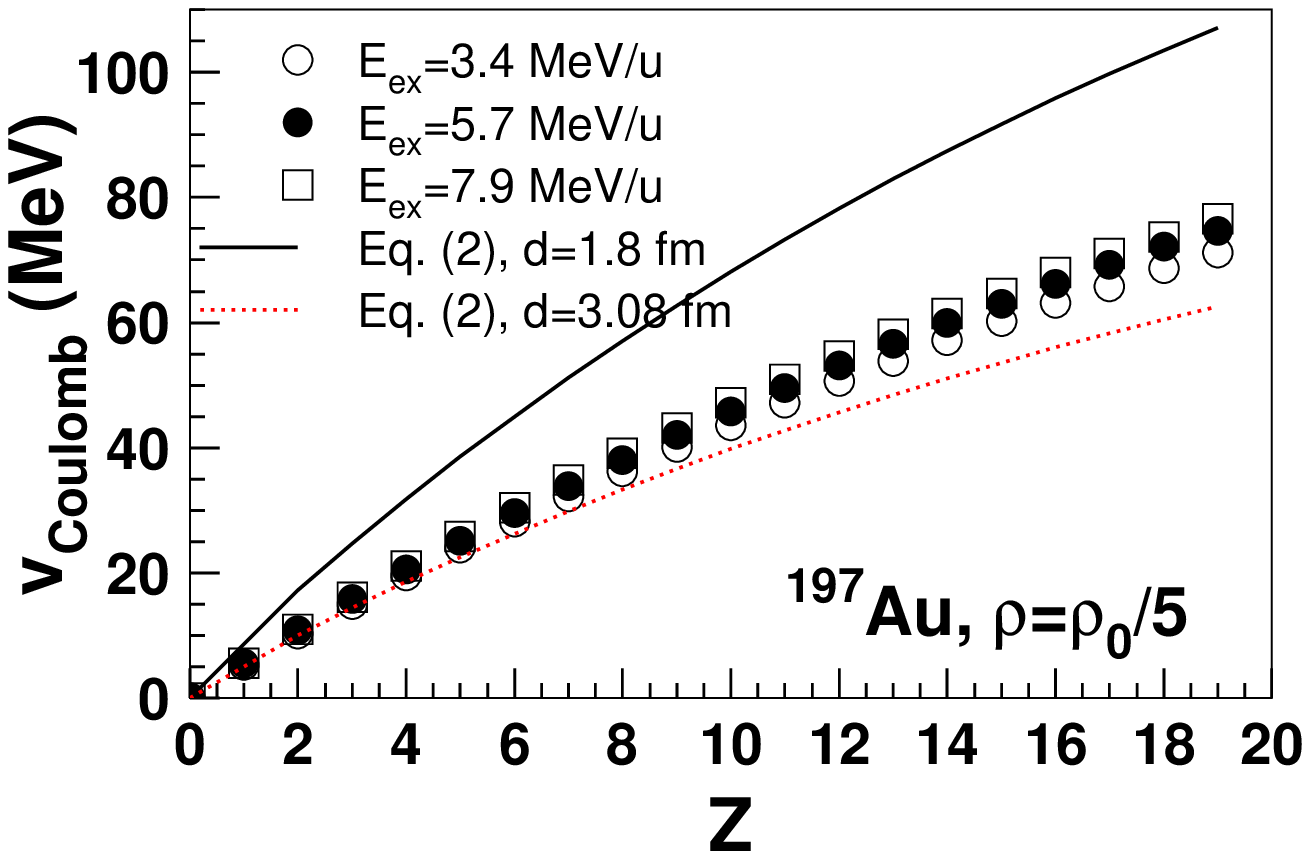}
  }
\caption{Left panel: Excitation energy dependence of average values of
total Coulomb energy, thermal energy and multiplicity of fragments with $Z \geq 3$
as a function of excitation energy
for $^{197}$Au multifragmenting nucleus at the freeze-out density, $\rho_0/5$
 as obtained by MMM;
Right panel: Average potential Coulomb energy experienced by a fragment
as a function of its charge for
$^{197}$Au, $\rho=\rho_0/5$ and $E_{ex}$=3.4 , 5.7 and 7.9 MeV/nucleon
as calculated with MMM.
The Coulomb barrier experienced by a fragment
calculated using Eq. \ref{eq:cb} 
is represented with 
lines assuming that both emitted fragment and residual
nucleus have normal nuclear
densities (d=1.8 fm) or densities equal to $\rho_0/5$ (d=3.08 fm), see text.}
\label{fig:coulomb_n=5}

\end{figure}
%%%%%%%%%%%%%%%%%%%%%%%%%%%%%%%%%%%%%%%%%%%%%%%%%%%%%%%%%%%%%%%%%%%%%%%%

To do that we shall use a microcanonical multifragmentation
model (MMM) \cite{mmm}
in order to study the excitation energy dependence
of the average Coulomb energy associated to the primary fragments
at freeze-out and the IMFs kinetic energy spectra. 
To keep the treatment as intuitive as possible we assimilate primary fragments
at break-up with spherical non-overlapping spheres placed in a spherical
container (the freeze-out
volume) and calculate Coulomb interaction using fragment-fragment interaction,
\begin{equation}
V_{\rm Coulomb}=1.44 \sum_{i<j} \frac{Z_i Z_j}{r_{ij}},
\label{eq:vcoulomb}
\end{equation}
where $Z_i$ denotes the charge of the fragment $i$, $r_{ij}$ stands for
the relative distance between two fragments and the sum runs over all
fragments of the given configuration such as to avoid double-counting.

For simplicity we assume that for all considered cases the size of the source
($^{197}$Au) and its break-up density are constant and modify only the
excitation energy.
As known from the early studies of multifragmentation, the increase of
excitation energy induces
an increase of both the degree of fragmentation and the thermal energy of the
system. A more advanced
fragmentation leads to a more uniform population of the available volume
and, consequently, 
to an increase of the total Coulomb energy of the system. However,
 by increasing the excitation energy, the number of fragments
at freeze-out increases much faster than
the associated total Coulomb energy which accounts for most of
the experimentally detected final kinetic
energy. Thus, one expects a reduced
increase of the average
Coulomb potential experienced
by any fragment due to the mean field generated by the other fragments.
These effects are illustrated in Fig. \ref{fig:coulomb_n=5}. 
In the left panel are plotted the average values of total Coulomb, thermal
energies and multiplicity of fragments with $Z \geq 3$
as a function of excitation energy
for $^{197}$Au at the freeze-out
density, $\rho_0/5$, while the right panel of Fig. \ref{fig:coulomb_n=5}
presents the average potential Coulomb energy 
experienced by a fragment as a function of its charge,

\begin{equation}
v_{\rm Coulomb}(Z)=\frac12 \cdot 1.44 \sum_{i (Z_i=Z)} Z_i \cdot \sum_j 
\frac{Z_j}{r_{ij}} \cdot \frac 1{y(Z)},
\label{eq:pot}
\end{equation}
where $y(Z)$ represents the average multiplicity of fragments with charge $Z$.
The obvious relation between the 
%total Coulomb energy 
total Coulomb energy $V_{\rm Coulomb}$
and the average Coulomb energies
experienced by different fragments is,
\begin{equation}
V_{\rm Coulomb}=\sum_{Z} y(Z) v_{\rm Coulomb} (Z).
\label{eq:rel}
\end{equation}

An increase of about 1.2 MeV/nucleon is obtained for the total Coulomb energy
when excitation energy moves from 3.4 to 7.9 MeV/nucleon (left panel) 
whereas, as the same time, a small increase of about 0.18 MeV/nucleon is
observed for example for Z=10 (right panel).
The estimation of the Coulomb contribution done using Eq.(\ref{eq:cb}) and d=3.08 fm
which corresponds to density $\rho_0/5$ is also shown on the right panel.
This value for d is
obtained taking d= 1.8 for normal density as suggested in Ref. \cite{viola}.
Estimations are indeed close to average values obtained
considering fragment-fragment interactions.
%%%%%%%%%%%%%%%%%%%%%%%%%%%%%%%%%%%%%%%%%%%%%%%%%%%%%%%%%%%%%%%%%%%%%%%%%%
\begin{figure}
\resizebox{0.7\textwidth}{!}{%
  \includegraphics{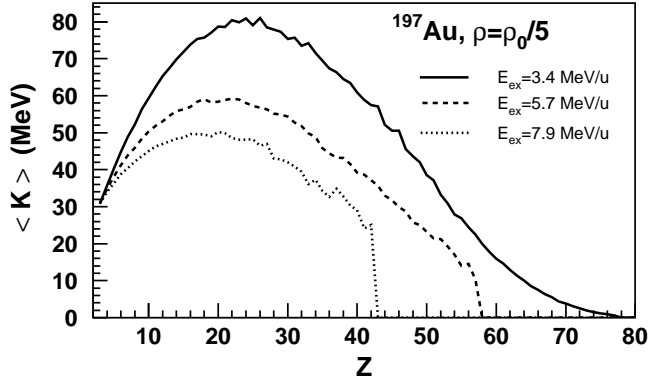}
  }
\caption{MMM predictions on break-up average kinetic energy as a function of fragment charge for $^{197}$Au
source at $\rho=\rho_0/5$ at $E_{ex}$=3.4, 5.7 and 7.9 MeV/nucleon.}
\label{fig:k_z_n=5}
\end{figure}
%%%%%%%%%%%%%%%%%%%%%%%%%%%%%%%%%%%%%%%%%%%%%%%%%%%%%%%%%%%%%%%%%%%%%%%%%%%%%
%%%%%%%%%%%%%%%%%%%%%%%%%%%%%%%%%%%%%%%%%%%%%%%%%%%%%%%%%%%%%%%%%%%%%%%%%%%%%%
\begin{figure}
\resizebox{0.95\textwidth}{!}{%
  \includegraphics{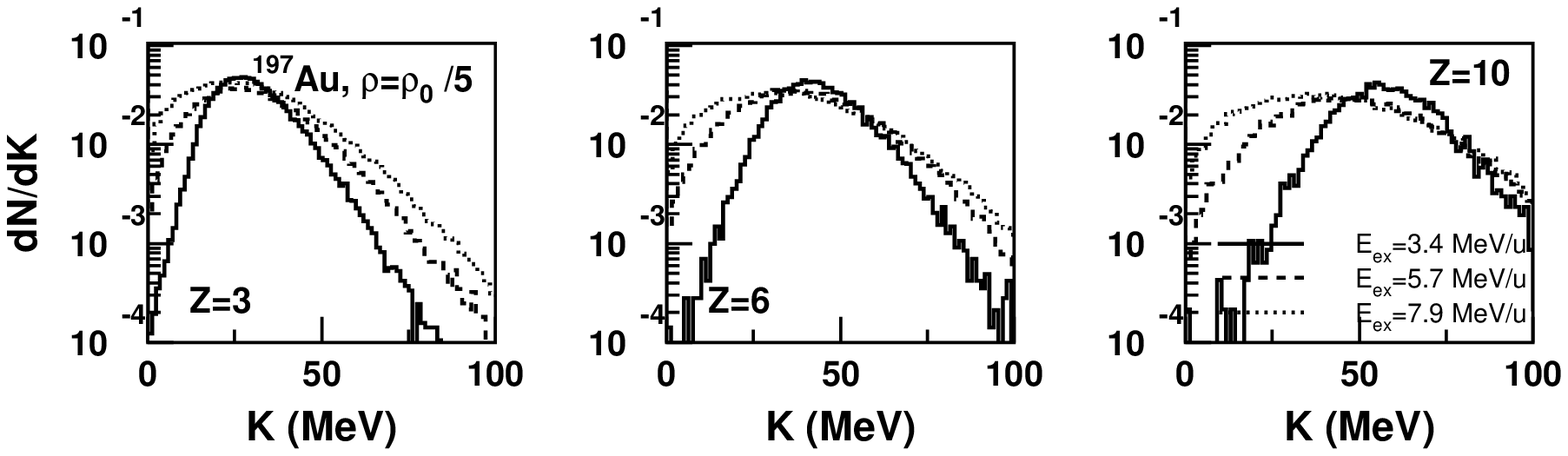}
  }
\resizebox{0.95\textwidth}{!}{%
  \includegraphics{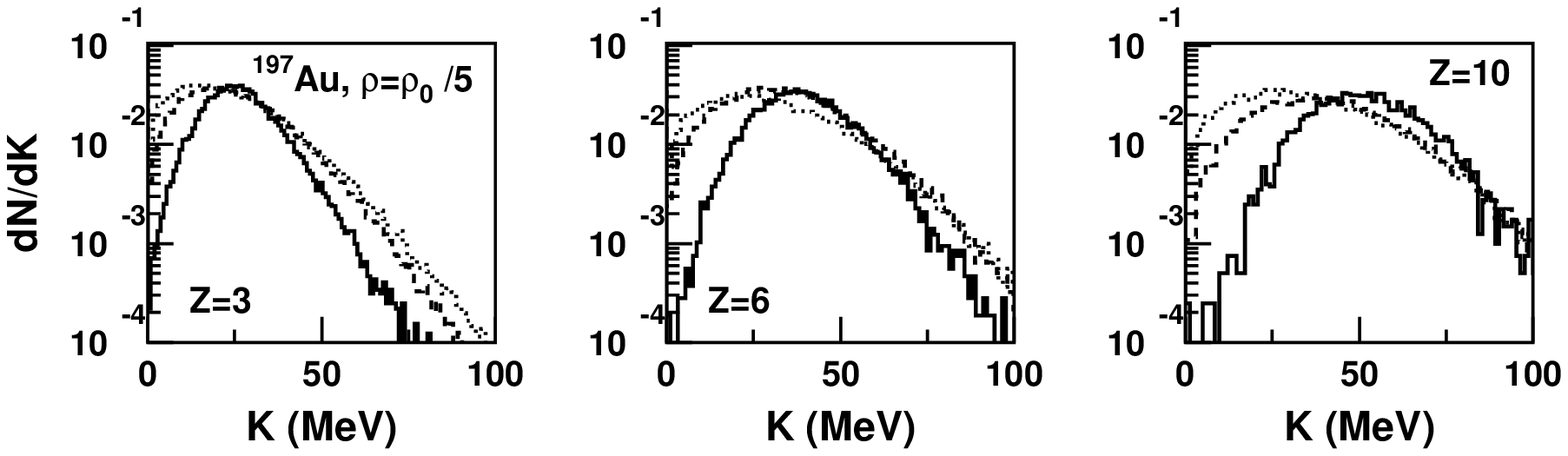}
  }

\caption{MMM predictions corresponding to break-up (upper panel) and asymptotic stage (lower panel)
kinetic energy spectra for different emitted intermediate mass fragments
resulted from the multifragmentation of $^{197}$Au at $\rho_0/5$
and different excitation energies.}
\label{fig:en_sp_n=5}

\end{figure}
%%%%%%%%%%%%%%%%%%%%%%%%%%%%%%%%%%%%%%%%%%%%%%%%%%%%%%%%%%%%%%%%%%%%%%%%%%%%%%%

Adding now the kinetic part of the thermal energy at freeze-out shared at
random between particles and fragments under constraints of
conservation laws, we can consider what is the effect of
increasing excitation energy on IMF average kinetic energies.
The mean kinetic energy distributions as a function of charge for the same
$^{197}$Au source and the same density $\rho_0/5$ at
$E_{ex}$=3.4, 5.7 and 7.9 MeV/nucleon are plotted in Fig. \ref{fig:k_z_n=5}.
At
%the
first glance, the behavior of $<K> (Z)$ distributions with increasing
source excitation is surprising in the sense that while both thermal and
Coulomb
energies increase, the fragment average kinetic energies decrease. This result
can
be understood having in mind the strong increase of fragment multiplicity which
leads to reduced kinetic energy per fragment. The narrowing of $<K> (Z)$
distributions
%is caused
is obviously caused
by the narrowing of
%$Y(Z)$
$y(Z)$ distributions once the excitation energy increases.

Clearly these results contradict the expectation of an increase with
temperature or excitation energy.
However they concern average quantities and not the peak
centroids. We can consider now the spectra.
As one may see in the upper panel of Fig. \ref{fig:en_sp_n=5} the modification
of the IMF kinetic energy
spectra is in qualitative agreement with the experimental data
cited in Ref. \cite{viola}: with
increasing $E_{ex}$ the centroids of the distribution move toward
smaller energies whereas their widths strongly increase. Since is known that
primary excited fragments undergo secondary 
emission, a natural question is whether or not this
process modifies the observed results. As one may see from
the lower panel of Fig. \ref{fig:en_sp_n=5} sequential evaporations
slightly diminish the IMF kinetic energies for a given Z, without
modifying the relative displacement of distributions corresponding to
different excitation energies.

In conclusion, using
a standard simultaneous multifragmentation model
we explained the experimentally evidenced evolution of the IMFs
kinetic energy spectra with increasing excitation energy as a consequence
of advanced system's fragmentation,
without any assumption regarding the modification of the break-up density.
To make our study as complete as possible, the behavior of both average
kinetic energy of IMFs and the IMFs kinetic energy spectra have been analyzed
for the freeze-out density range usually addressed by statistical
multifragmentation models, namely $\rho_0/7$ to $\rho_0/3$. The obtained
results are qualitatively the same as the above presented results
corresponding to $\rho_0/5$.
This study suggests that an alternative explanation as compared to
the conclusions of Ref. \cite{viola}
%connected to the description of multifragmentation
%as a sequential process can be proposed.
can be proposed, which is connected to a different description of
multifragmentation.
Answering the important question on what is the break-up density dependence
on the excitation energy needs more consideration.

\end{document}